\documentclass{sf2a-conf}
\usepackage{graphicx}
\usepackage{natbib}

\begin{document}

\TitreGlobal{SF2A 2008}

\title{Outflows, Bubbles, and the Role of the Radio Jet:
Direct Evidence for AGN Feedback at z$\sim$2}
\author{Nesvadba, N.,~P.,~H.$^{1,}$}
\address{GEPI, Observatoire de Paris, CNRS,
  Universite Denis Diderot; 5, Place Jules Janssen, 92190 Meudon, France} 
\address{Marie-Curie Fellow}
\author{Lehnert,~M.~D.$^1$}

\setcounter{page}{237}

\index{Nesvadba, N.}
\index{Lehnert, M.}

\maketitle
\begin{abstract} 
To accommodate the seemingly "anti-hierarchical" properties of galaxies near
the upper end of the mass function within our hierarchical paradigm, current
models of galaxy evolution postulate a phase of vigorous AGN feedback at high
redshift, which effectively terminates star formation by quenching the supply
of cold gas. Using the SINFONI IFU on the VLT, we identified kpc-sized
outflows of ionized gas in z$\sim$2$-$3 radio galaxies, which have the
expected signatures of being powerful AGN-driven winds with the potential of
terminating star formation in the massive host galaxies. The bipolar outflows
contain up to few$\times 10^{10}$ M$_{\odot}$ in ionized gas with velocities
near the escape velocity of a massive galaxy. Kinetic energies are equivalent
to $\sim 0.2$\% of the rest mass of the supermassive black hole. 
We discuss the results of this on-going study and the global impact of the
observed outflows.
\end{abstract}
%
\section{The role of AGN feedback for galaxy evolution in the early universe} 
AGN feedback is now a critical element of state-of-the-art models of galaxy
evolution tailored to solve some of the outstanding issues at the upper end of
the galaxy mass function. Observationally, a picture emerges where AGN
feedback is most likely related to the mechanical energy output of the
synchrotron emitting, relativistic plasma ejected during the radio-loud phases
of AGN activity: Giant cavities in the hot, X-ray emitting halos of massive
galaxy clusters filled with radio plasma are robust evidence for AGN feedback
heating the gas on scales of massive galaxy clusters (e.g., McNamara \&
Nulsen, 2007). Best et al. (2006) analyzed a large sample of early-type
galaxies from the SDSS catalog with FIRST and NVSS radio data and found that
heating by the radio source may well balance gas cooling over 2 orders of
magnitude in radio power and in stellar mass.

However, since most of the growth of massive galaxies was completed during the
first few Gyrs after the Big Bang, observations at low redshift can only
provide evidence that AGN feedback is able to {\it maintain} the hot,
hydrostatic halos of massive early-type galaxies ({\it ``maintenance
mode''}). If we want to observe directly whether AGN feedback indeed quenched
star formation and terminated galaxy growth in the early universe ({\it
``quenching mode''}), we have to search at high redshift. With this goal, we
started a detailed analysis of the rest-frame optical line emission in
powerful, z$\sim$2$-$3 radio galaxies with integral field spectroscopy, where
we may plausibly expect the strongest signatures of AGN-driven winds.

\section{Powerful radio galaxies at z$\sim$ 2$-$3: Dying starbursts in 
the most massive galaxies?} 
The observed properties of HzRGs suggest they may be ideal candidates to
search for strong, AGN-driven winds: They have large stellar (Seymour et
al. 2007) and dynamical (Nesvadba et al. 2007a) masses of $\sim$$10^{11-12}$
M$_{\odot}$ and reside in significant overdensities of galaxies suggesting
particularly massive underlying dark-matter halos (e.g., Venemans et
al. 2007). Large molecular gas masses in some sources (e.g., Papadopoulos et
al. 2000) and submillimeter observations suggest that some HzRGs at redshifts
z$\ge 3-4$ are dust-enshrouded, strongly star-forming galaxies with FIR
luminosities in the ULIRG regime. Interestingly, the fraction of
submillimeter-bright HzRGs shows a rapid decline from $>$50\% at z$>$2.5 to
$\le$15\% at z$<$2.5 (Reuland et al. 2004).  This suggests that HzRGs may be
particularly massive galaxies near the end of their phase of active star
formation. They also host particularly powerful AGN. Thus, they are good
candidates to search for the kinematic signatures of AGN-driven winds.

\section{Observational evidence for AGN-driven winds in 
z$\sim$2$-$3 radio galaxies} 
To directly investigate whether HzRGs may be the sites of powerful, AGN driven
winds, we collected a sample of HzRGs at redshifts z$\sim$2$-$3 with
rest-frame optical near-infrared spectral imaging obtained with SINFONI on the
VLT. Including scheduled observations, our total sample will consist of 29
galaxies spanning wide ranges in radio power and radio size. We also include
galaxies with compact, and probably young, radio sources. We will in the
following concentrate on the analysis of a first subsample of 6 galaxies, 4
with extended jets with radii between 10 and 50 kpc, 2 with more compact radio
sources $<$10 kpc in radius. For details see Nesvadba et al. (2006, 2007a,
2008).

\begin{figure}
   \centering
   \includegraphics[width=0.37\textwidth]{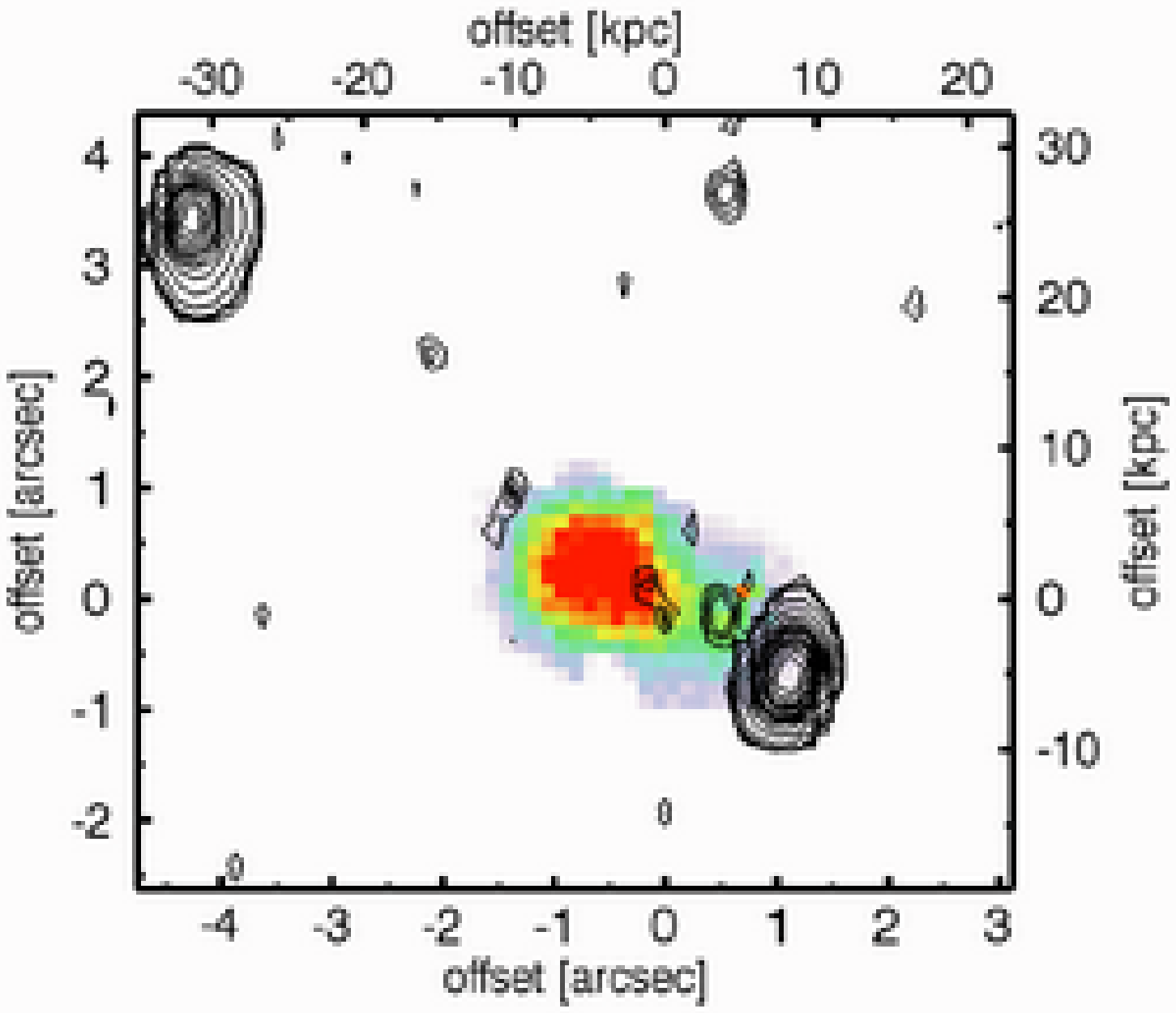} 
   \includegraphics[width=0.32\textwidth]{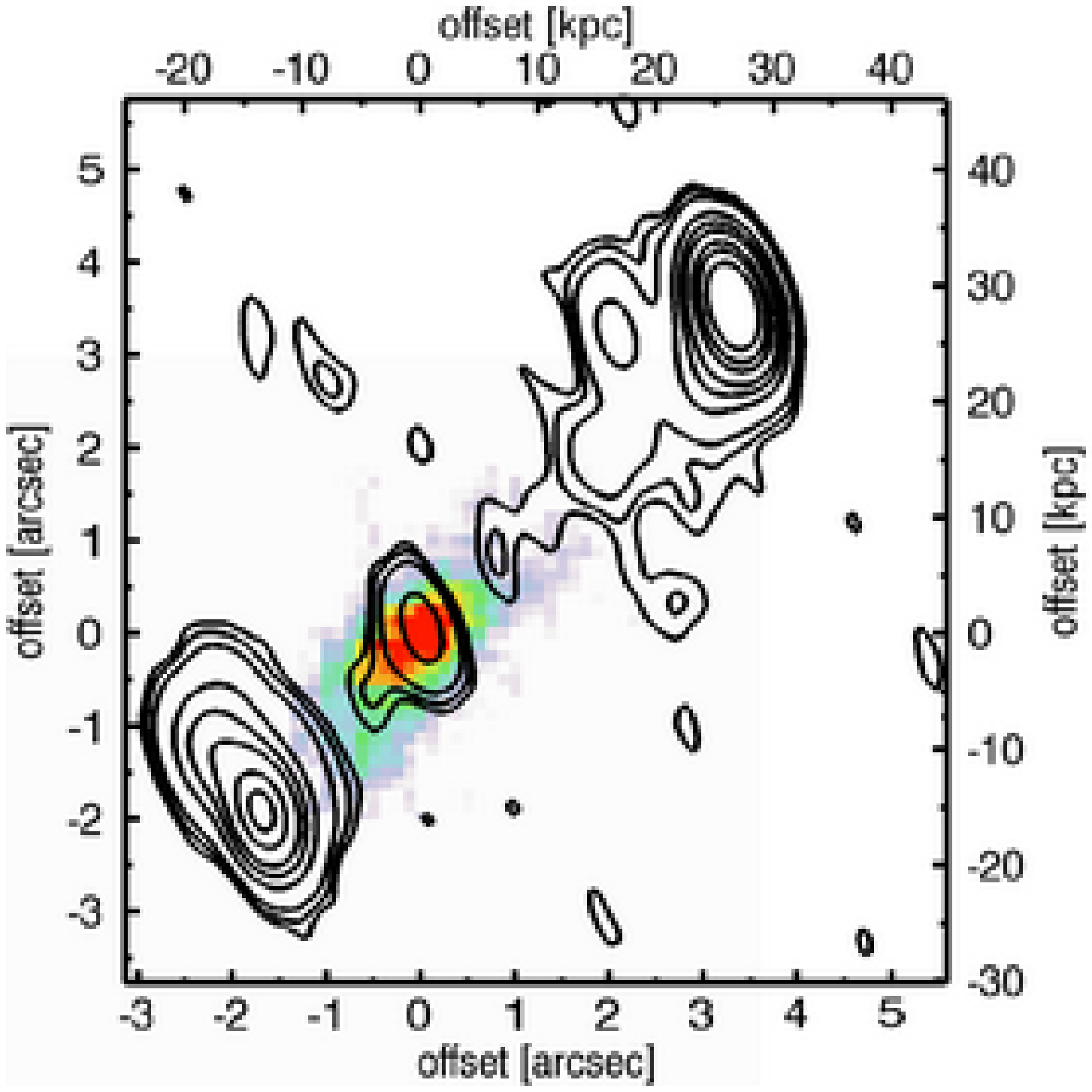}
   \includegraphics[width=0.29\textwidth]{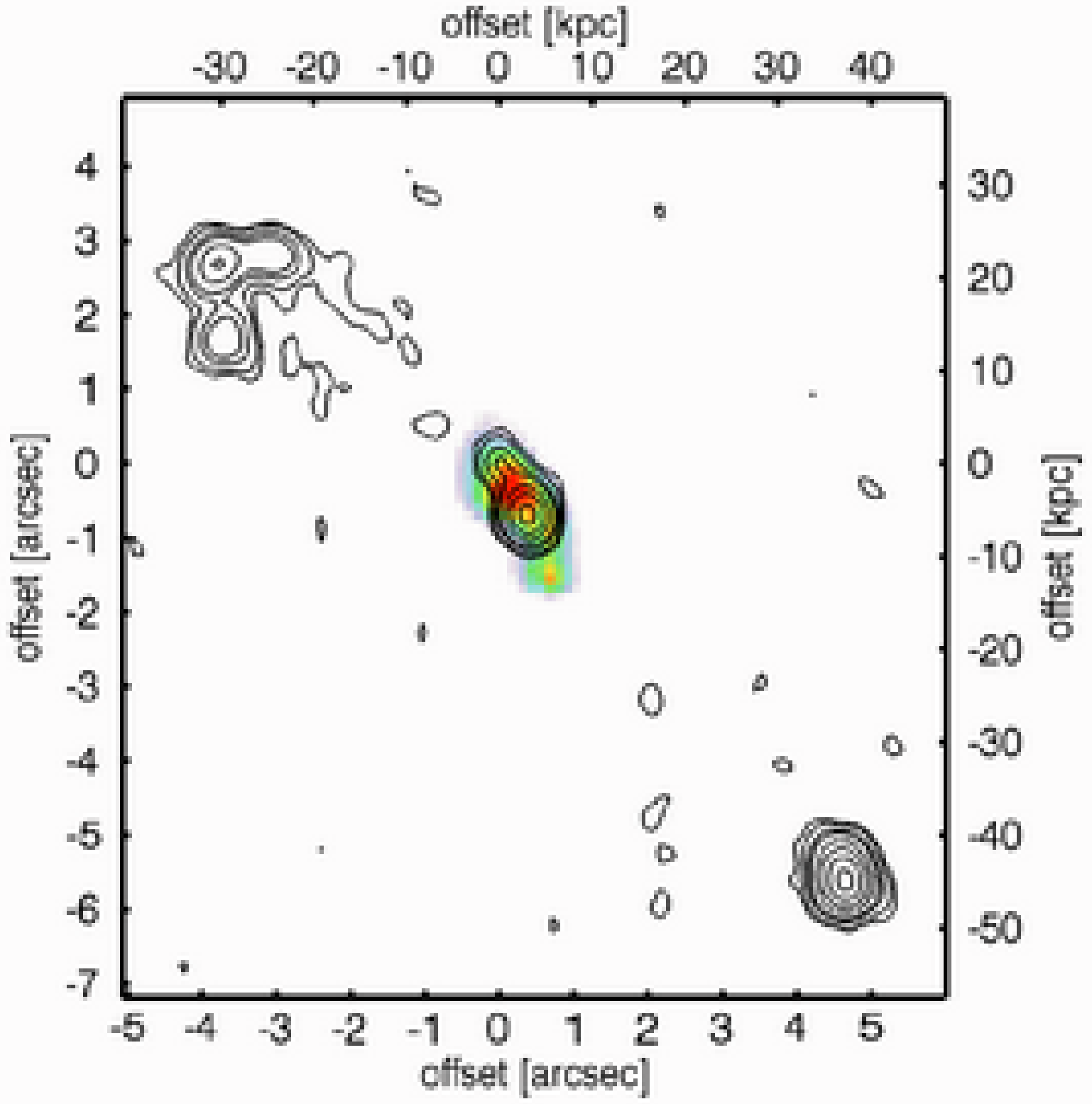}
      \caption{{\it (left to right:)} [OIII]$\lambda$5007 emission line
morphologies of MRC0316-257 at z$=$3.1, MRC0406-244 at z$=$2.4 and TXS0828+193
at z$=$2.6. Contours indicate the line-free continuum morphology for
MRC0406-244 and TXS0828+193, and the 4.8 GHz radio core for MRC0316-257, where
we did not detect the continuum. 
} 
\label{nesvadba2_fig1}
\end{figure}

\subsection{Continuum and emission line morphologies}
Using an integral-field spectrograph, we were able to extract continuum-free
line images as well as line-free continuum images from our three-dimensional
data cubes (Fig.~1). We find that in all cases, the continuum emission is
relatively compact, but spatially resolved in some cases, with half-light
radii $\le 5$ kpc. Radio-loud AGN activity is often related to an on-going
merger. However, we only identify one continuum knot per galaxy. For the
merger scenario, this may suggest an advanced stage where the galaxies are
seperated by less than the $\sim$ 4 kpc spatial resolution of our
data. Alternatively, since SINFONI is relatively inefficient in detecting low
surface-brightness continuum emission, nuclear activity may have been
triggered by other processes like minor mergers or perhaps cooling flows in
cluster environments. 

The extended, distorted morphologies of HzRGs with extended jets seen in
broad-band imaging are mostly due line contamination, originating from
emission line regions that extend over several 10s of kpc, and are
significantly larger than the continuum (Nesvadba et al. 2008), but extend to
smaller radial distances than the radio lobes. The same is found from
Ly$\alpha$ longslit spectroscopy (e.g., Villar-Martin 2003). Overall,
different emission lines in the same galaxy show similar morphologies. In the
galaxies with {\it compact} radio sources, the line emission appears also
compact. This may suggest a causal relationship between the advance of the jet
and the extent of the high surface brightness emission line gas. 

\subsection{Kinematics, outflow energies, and physical properties of the
 ionized gas} 
\label{ssec:kinematics}
We fitted spectra extracted from individual spatial resolution elements to
construct two-dimensional maps of the relative velocities and line widths
(Fig. 2). Typically, the velocity maps show two bubbles with relatively
homogeneous internal velocity, and projected velocities relative to each other
of 700$-$1000 km s$^{-1}$, reminiscent of back-to-back outflows extending from
near the radio core. MRC1138-262 has a more complex structure with at least 3
bubbles. Line widths are generally large, indicating strong turbulence, with
typical FWHMs $\sim$500$-$1200 km s$^{-1}$. Areas with wider lines may be due
to partial overlap between bubbles.

Filamentary morphologies and low gas filling factors suggest that the
UV-optical line emission may originate from clouds of cold gas that are being
swept up by an expanding hot medium, most likely related to the
overpressurized 'cocoon' of gas heated by the radio jet. In such a scenario
the velocity of the clouds may yield an estimate of the kinetic energy
injection rate necessary to accelerate the gas to the observed velocities of
up to $\sim 10^{45}$ erg s$^{-1}$ (Nesvadba et al. 2006). The size and
velocities of the outflow suggest dynamical timescales of few $\times 10^7$
yrs. Maintaining the observed outflows over such timescales requires total
energy injections of $\sim 10^{60}$ erg. This is in the range of what is
observed for AGN driven bubbles in massive clusters at low redshift (e.g.,
McNamara \& Nulsen, 2006, and references therein). The observed velocities and
kinetic energies are also in the range of escape velocities and binding
energies expected for galaxies with masses of few $\times 10^{11}$ M$_{\odot}$
(Nesvadba et al. 2006). This may suggest that much of the gas participating in
the outflows may ultimately be unbound from the underlying gravitational
potential.

\begin{figure}[t]
   \centering
   \includegraphics[height=0.27\textwidth]{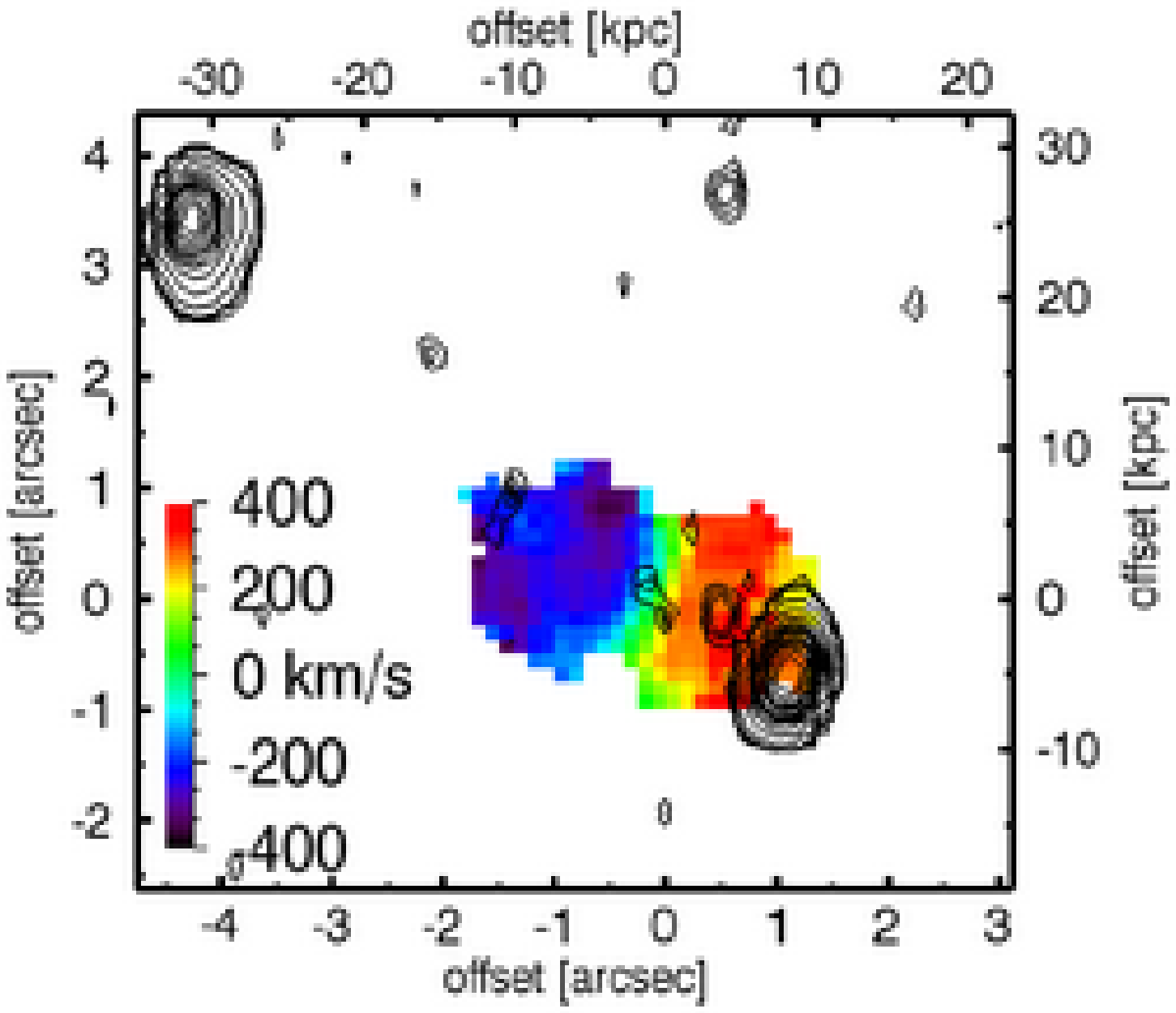}
   \includegraphics[height=0.25\textwidth]{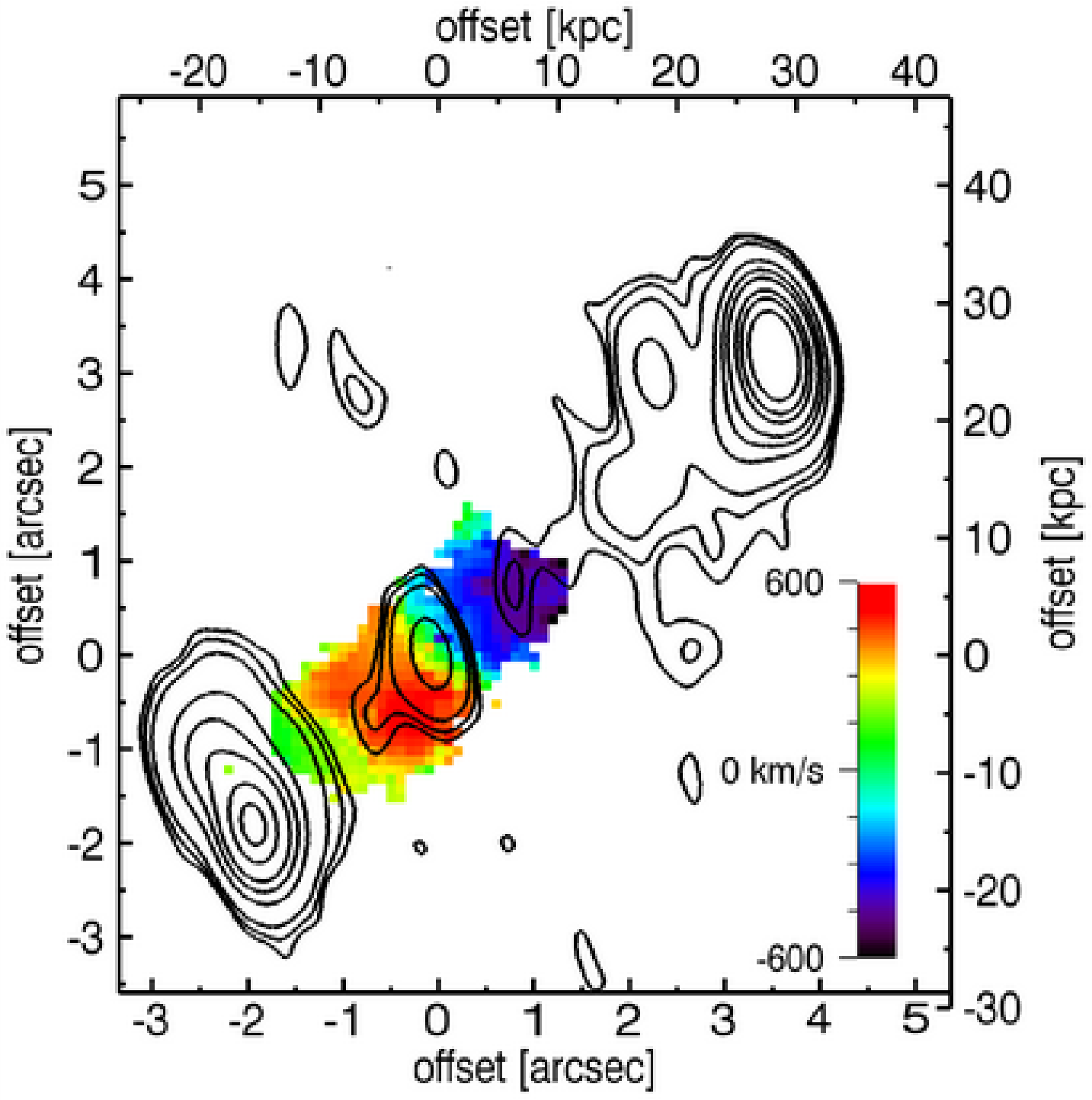} 
   \includegraphics[height=0.25\textwidth]{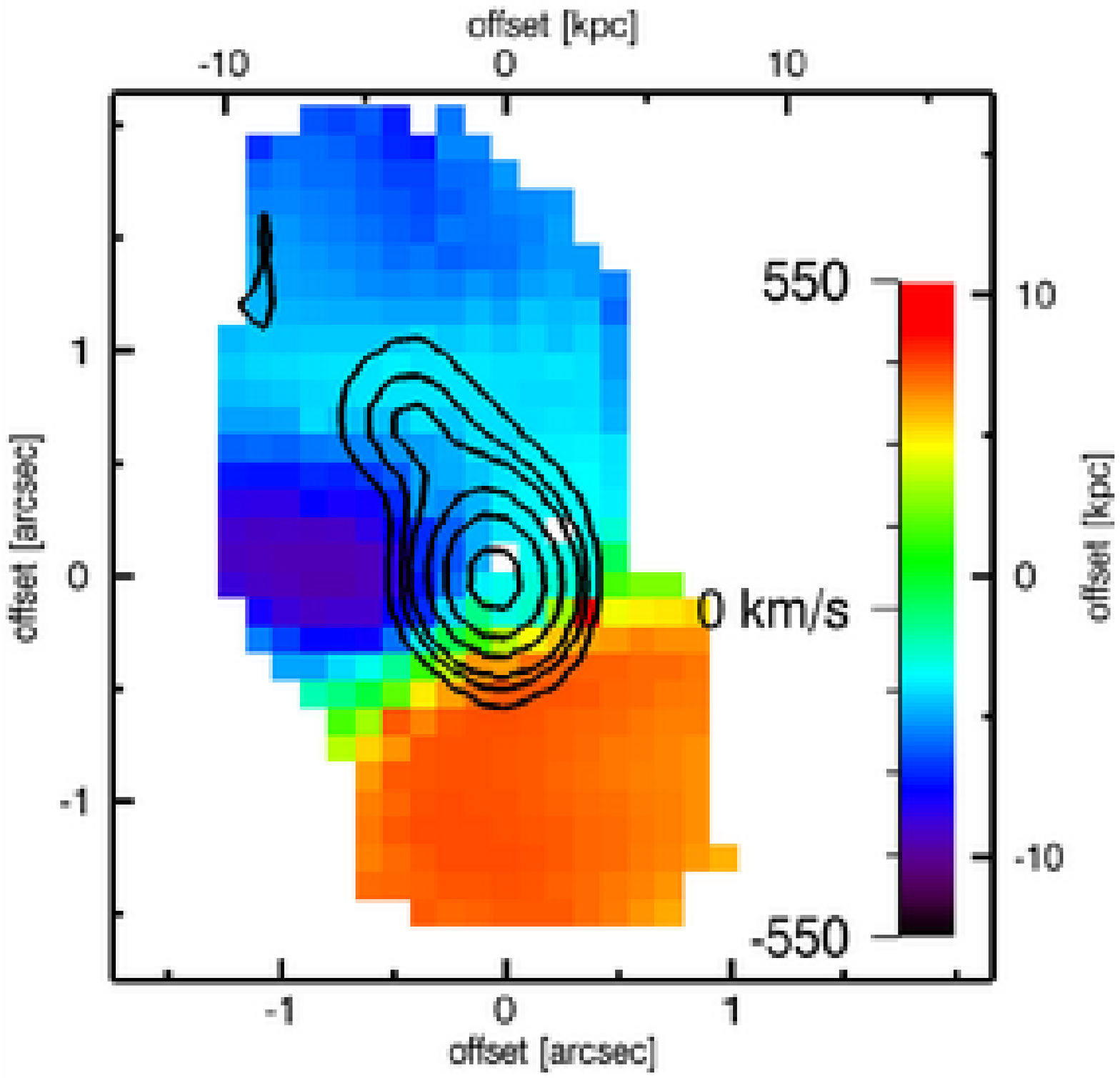}\\
   \includegraphics[height=0.27\textwidth]{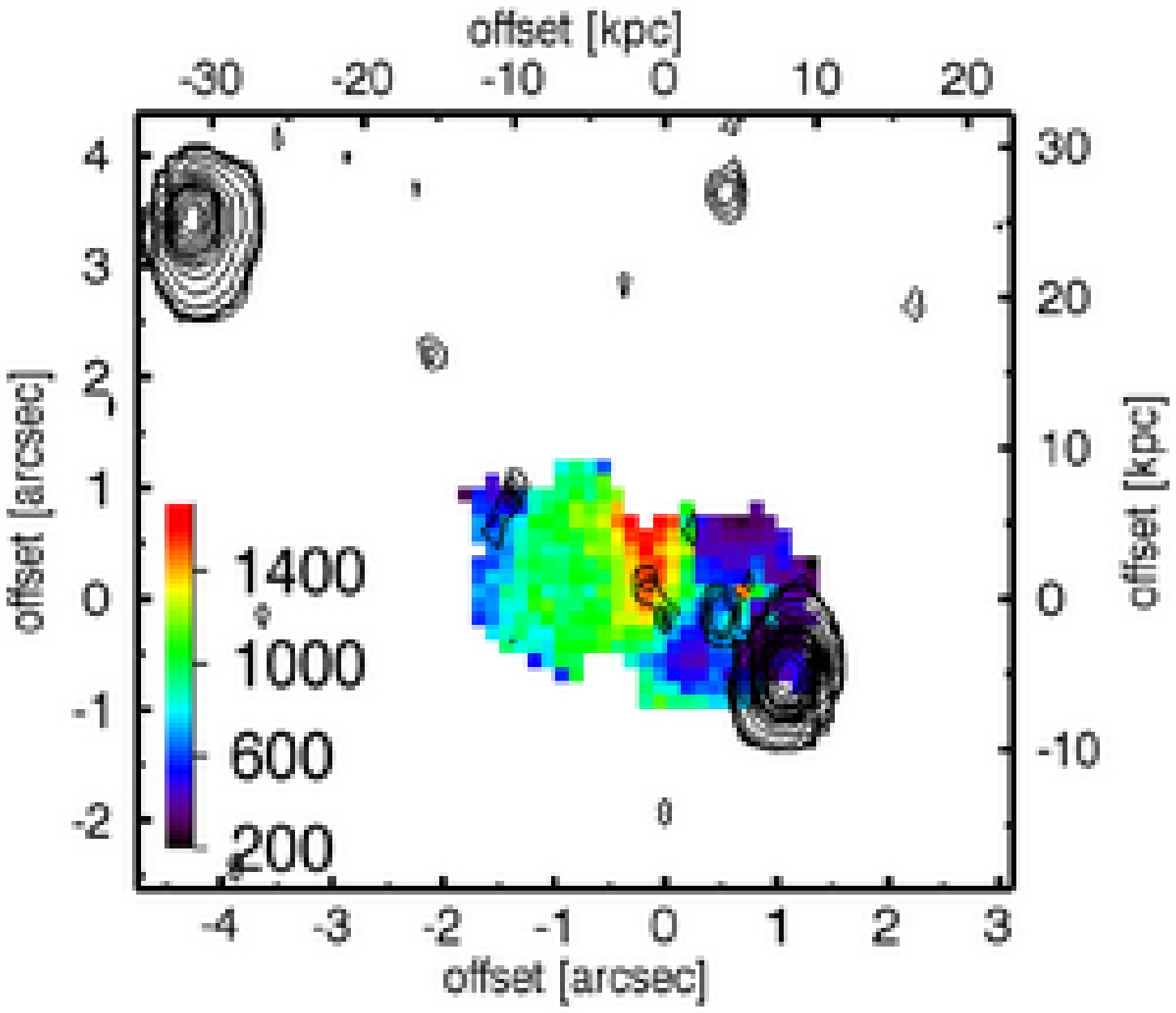}
   \includegraphics[height=0.25\textwidth]{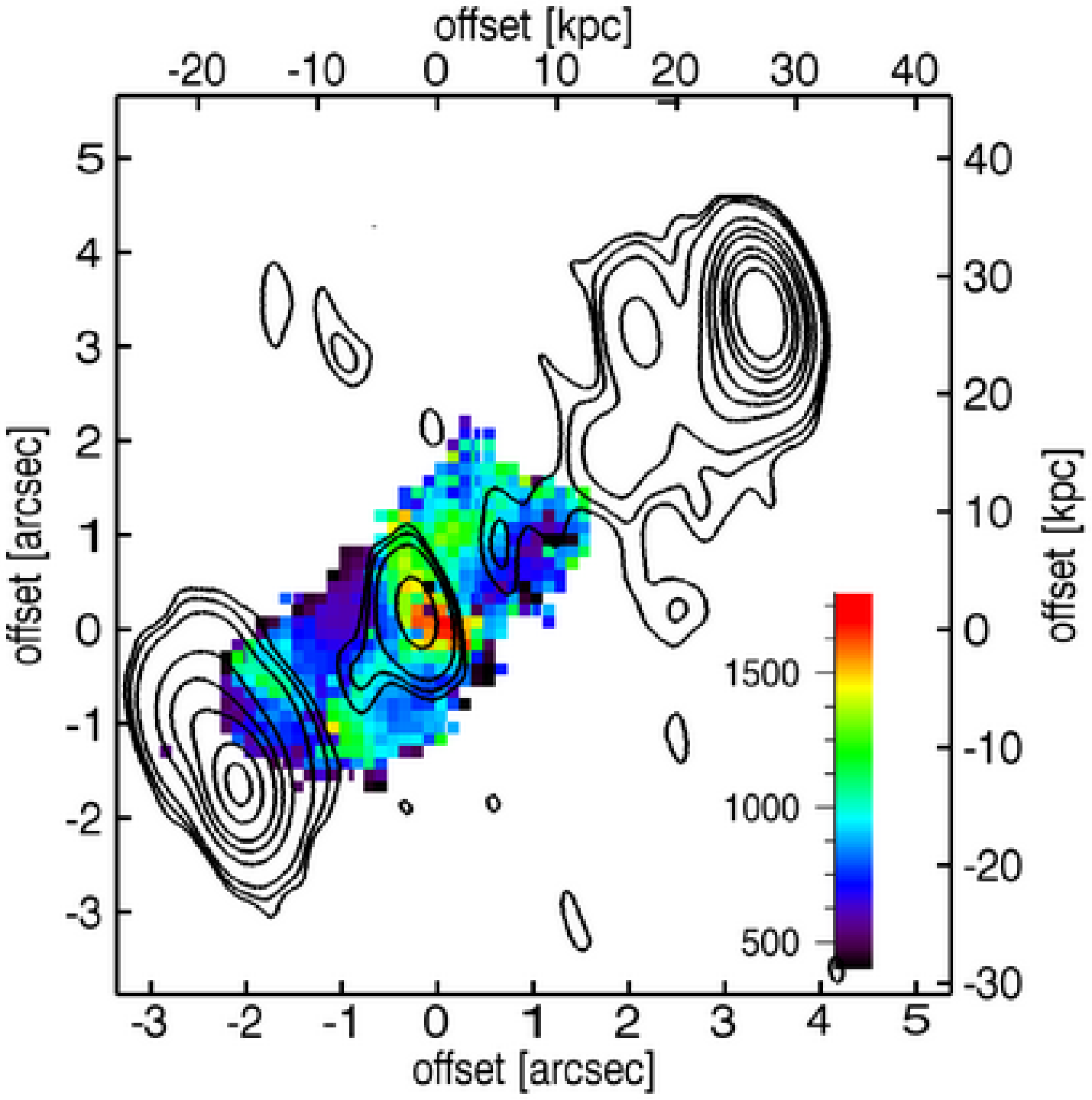}
   \includegraphics[height=0.25\textwidth]{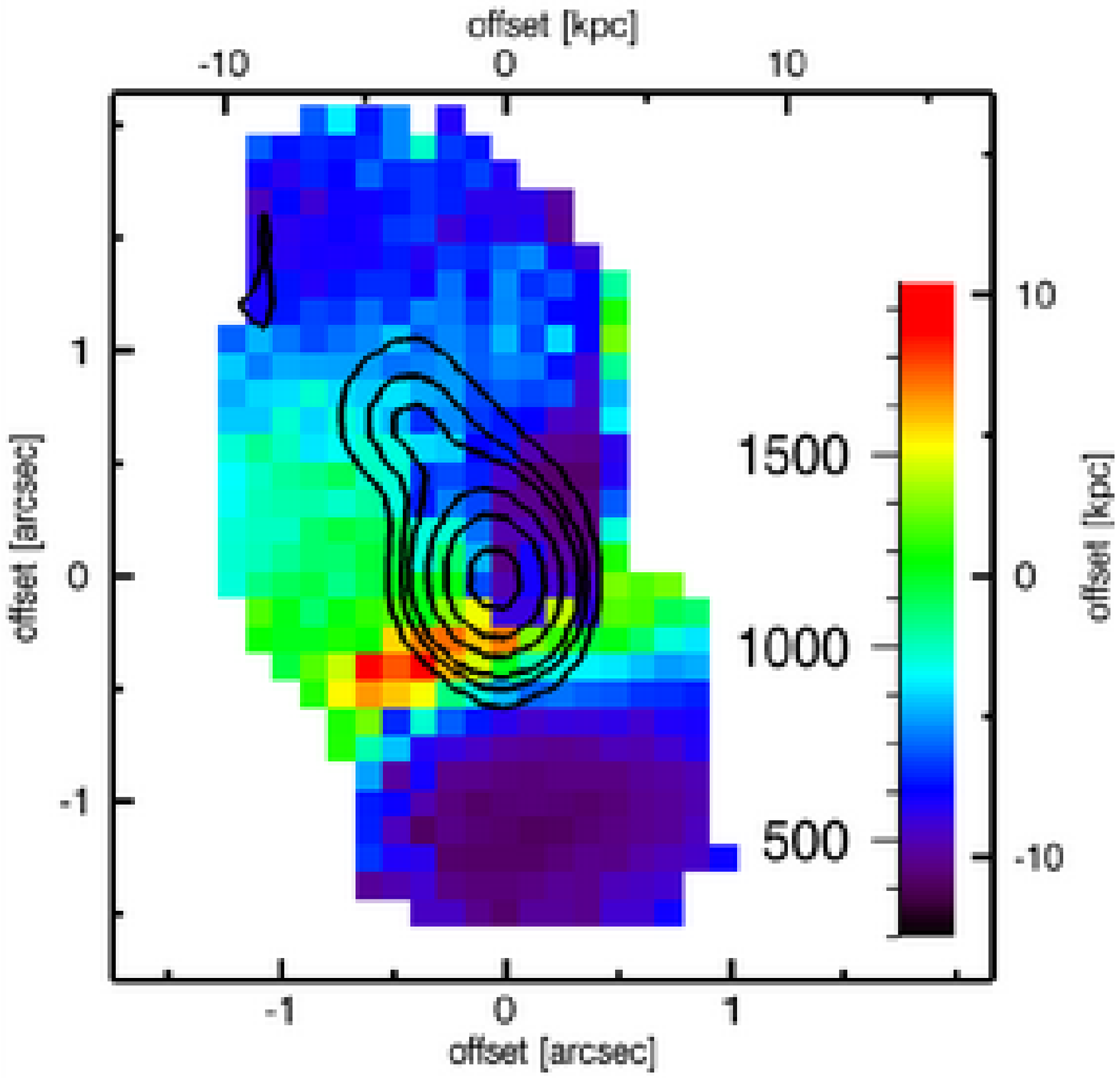}
      \caption{{\it top, left to right:} Maps of relative velocities (in km
        s$^{-1}$) for MRC0316-257 at z$=$3.1, MRC0406-244 at z$=$2.4, and
        TXS0828+193 at z$=$2.6. {\it bottom, left to right}: Maps of FWHMs (in
        km s$^{-1}$) for the same galaxies. Contours show the jet
        morphologies. For 
        TXS0828+193, the lobes are outside of the area shown.}
       \label{nesvadba2_fig2}
\end{figure}

\subsection{Molecular and ionized gas budgets}
Having measured H$\alpha$ line fluxes, we are able to roughly estimate ionized
gas masses assuming case B recombination (see Nesvadba et al. 2008 for
details). For galaxies where we also measured H$\beta$, we correct for
extinction of A$_V\sim$1$-$4 mag and find ionized gas masses of up to few
$\times$ $10^{10}$ M$_{\odot}$. (Without the correction, estimates are few
$\times 10^9$ M$_{\odot}$, Nesvadba et al. 2008.) This exceeds the amount of
ionized gas found in any other high-redshift galaxy population by several
orders of magnitudes, including galaxies with starburst-driven winds. Nesvadba
et al. (2007b) investigated a spatially-resolved, starburst driven wind in a
strongly star-forming submillimeter-selected galaxy at z$\sim$2.6 with of
order few$\times$$10^6$ M$_{\odot}$ in ionized gas in the wind. Compact radio
galaxies have lower entrained gas masses, but in the range of what would be
expected for less evolved outflows with similar entrainment rates as the
galaxies with large radio lobes (Nesvadba et al. 2007a).

Molecular gas masses in strongly star-forming galaxies at high redshift are
also typically in the range of few $\times 10^{10}$ M$_{\odot}$ (e.g., Neri et
al. 2003), and are a necessary prerequisite to fuel the observed starbursts
with star formation rates of few 100 M$_{\odot}$ yr$^{-1}$. However, not all
HzRGs have been detected in CO. TXS0828+193 specifically, which is part of our
sample, appears to have less than $\sim 10^{10}$ M$_{\odot}$ in molecular gas
(Nesvadba et al., in prep.). This illustrates that the AGN winds may affect a
significant fraction of the overall interstellar medium of strongly
star-forming, massive galaxies in the early universe. Since the velocities are
near the expected escape velocity of a massive galaxy and underlying
dark-matter halo (\S\ref{ssec:kinematics}), much of this gas may actually
escape.

\section{Global impact of AGN driven winds}
Four out of four HzRGs with extended radio jets show evidence for outflows
with with kinetic energies of up to 10$^{60}$ erg over dynamical timescales of
$10^7$ yrs, and the preliminary analysis of our full sample suggests that this
is far from being unusual. Nesvadba et al. (2006, 2008) estimate that the
outflow energies correspond to $\sim 10$\% of the jet kinetic luminosity. If
this coupling efficiency between jet and interstellar medium is typical for
HzRGs with similarly powerful radio sources, then the redshift-dependent
luminosity function of Willott et al. (2001) suggests that at redshifts
z$\sim$ 1$-$3, AGN-winds release an overall energy density of about $10^{57}$
erg s$^{-1}$ Mpc$^{-3}$. Some of this energy release may contribute to
heating and increasing the entropy in extra-galactic gas surrounding the HzRG,
and to enhance gas stripping in satellite galaxies, so that subsequent merging
with satellites will be relatively dissipationless. This may later contribute to
preserving the low content in cold gas and old, luminosity weighted ages of
the highly metal-enriched stellar population in massive galaxies to the
present day, in spite of possible continuous accretion of satellite galaxies
over cosmologically significant periods (Nesvadba et al. 2008).

If the outflows are related to the nuclear activity, then the ultimate energy
source powering the outflow is accretion onto the supermassive black hole
in the center of the galaxy. Thus, models of galaxy evolution often
parameterize the efficency of AGN feedback by the energy equivalent of the rest
mass of the black hole. Since we have reason to believe that HzRGs
approximately fall onto the low-redshift M-$\sigma$ relationship between the
mass of the supermassive black hole and velocity dispersion of the host, we
can use the stellar mass estimates of Seymour et al. (2007) to roughly
estimate the black hole mass of our targets. We find that of order 0.1\% of
the energy equivalent of the black hole mass in HzRGs is being released in
kinetic energy of the outflows. A similar estimate based on the global energy
density released by powerful radio galaxies estimated above, and the local
black hole mass density yields a very similar result, $\sim 0.2$\%. This is
very close to what is assumed in galaxy evolution models (e.g., Di Matteo et
al. 2005), and highlights the likely importance of the observed outflows on
galaxy evolution.

\end{document}